\documentclass[12pt,preprint]{aastex}

\slugcomment{Not to appear in Nonlearned J., 45.}

\shorttitle{The {\it Swift} Discovery of SGR X-ray Afterglows}
\shortauthors{Nakagawa et al.}

\begin{document}


\title{The {\it Swift} Discovery of X-ray Afterglows Accompanying Short Bursts from SGR\,1900$+$14}


\author{
Y. E. Nakagawa,\altaffilmark{1,2}
T. Sakamoto,\altaffilmark{3,4}
G. Sato,\altaffilmark{3}
N. Gehrels,\altaffilmark{3}
K. Hurley,\altaffilmark{5}
AND
D. M. Palmer,\altaffilmark{6}
}


\altaffiltext{1}{Graduate School of Science and Engineering, Aoyama Gakuin University, Sagamihara, Kanagawa 229-8558, Japan}
\altaffiltext{2}{Institute of Physical and Chemical Research (RIKEN), 2-1 Hirosawa, Wako, Saitama 351-0198, Japan}
\altaffiltext{3}{NASA Goddard Space Flight Center, Greenbelt, MD 20771}
\altaffiltext{4}{University of Maryland, Baltimore County, 1000 Hilltop Circle, Baltimore, MD 21250}
\altaffiltext{5}{Space Sciences Laboratory, 7 Gauss Way, University of California, Berkeley, CA 94720-7450}
\altaffiltext{6}{Los Alamos National Laboratory, P.O.Box 1663, Los Alamos, NM 87545}


\begin{abstract}
The discovery of X-ray afterglows accompanying two short bursts from SGR\,1900$+$14
is presented.
The afterglow luminosities at the end of each observation are lower by 30-50\%
than their initial luminosities, and decay with power law
indices $p \sim $0.2-0.4.
Their initial
bolometric luminosities are $L \sim 10^{34}$-$10^{35}$\,erg\,s$^{-1}$.
We discuss analogies and differences between the X-ray afterglows
of SGR short bursts and short gamma-ray bursts.
\end{abstract}





\section{Introduction}
Soft gamma repeaters (SGRs) are intriguing sources of
very energetic (super-Eddington luminosity) high energy bursts.
They exhibit repetitive, sporadic bursting activity
with typical burst durations of $\sim100$\,ms \citep[e.g.,][]{woods2006}.
Four have been identified as definite SGRs, three are candidate SGRs,
and the X-ray source AX\,J1818.8$-$1559 is either a candidate SGR
or an anomalous X-ray pulsar (AXP) \citep{mereghetti2007, nakagawa2007b}.
Quiescent X-ray emission has been observed from SGRs
with a flux which exceeds the Eddington luminosity
\citep{nakagawa2007b}.
The energy reservoir for the bursts and the steady emission
is generally believed to be magnetic energy dissipation
in the framework of the magnetar model
\citep[e.g.,][]{duncan1992}.

Among the SGR bursts, there is a minority population with
a few seconds duration known as intermediate bursts
\citep[e.g.,][]{olive2004}.
Some of the intermediate duration bursts from SGR\,1900$+$14 have
an X-ray tail
with a duration of a few thousand seconds that is interpreted
as a part of the burst itself \citep{lenters2003}.
More rarely, giant flares occur with an initial short, intense spike,
several hundred milliseconds long, followed by a long pulsating tail lasting a few hundred seconds;
these are the most exotic magnetar phenomena.
A large flare from SGR\,1900$+$14 on 2001 April 18
displayed an X-ray afterglow
lasting about 11 days \citep{feroci2003}.
Radio afterglows were observed after the
giant flares from SGR\,1900$+$14 on 1998 August 27 and from SGR\,1806$-$20
on 2004 December 27 \citep{frail1999, cameron2005}.
Indeed it is possible that some short-duration cosmic gamma-ray bursts (GRBs)
could actually be extragalactic giant magnetar flares \citep{hurley2005}.
Radio, optical, and/or X-ray afterglows are a common phenomenon for
the long-duration GRBs (typical durations of 2 seconds or more),
but they have only been observed for 13 of 20 GRBs of duration less than 2 seconds
(hereafter short GRBs).
In addition, a flux
increase and a slow decay of the quiescent emission
($F \propto t^{-0.3}$ over 110 days, where $F$ is the
flux and $t$ is the time since the burst)
has been observed after a short burst from the
AXP\,CXOU\,J164710.2-455216
\citep[e.g.,][]{israel2007, nakagawa2007b}.
A flux increase and a decay having two different components
($F \propto t^{-4.8\pm0.5}$ for $t < 0.5$\,day and
$F \propto t^{-0.22\pm0.01}$ for $t \gtrsim 0.5$\,day)
has been observed in the quiescent emission
after an outburst from AXP\,1E\,2259$+$586 in 2002 \citep{woods2004}.

Many bursts from the known SGRs have been detected by {\it Swift} \citep{gehrels2004}
because of the very wide field-of-view and excellent sensitivity
of the Burst Alert Telescope \citep[BAT;][]{barthelmy2005}.
Thanks to the prompt, precise localization by the BAT,
and the fast slew capability of {\it Swift}, X-ray follow-up observations
by the X-ray Telescope \citep[XRT;][]{burrows2005}
have taken place soon after short bursts from SGR\,1900$+$14.
In this paper, we report the {\it Swift} discovery of X-ray afterglows
from three of these short bursts.
We present spectral analyses of the bursts and the X-ray afterglow
using BAT and XRT data.
We discuss analogies and differences between the short bursts from the SGRs
and the short cosmic GRBs.
Despite many satellite and ground-based telescope observations,
the distance to SGR\,1900$+$14 still remains very uncertain.
In this paper, the distance is assumed to be 10\,kpc \citep{hurley1999, vrba2000}.

\section{Data Analysis}

\subsection{Observations}
\begin{figure}
\epsscale{1.}
\plotone{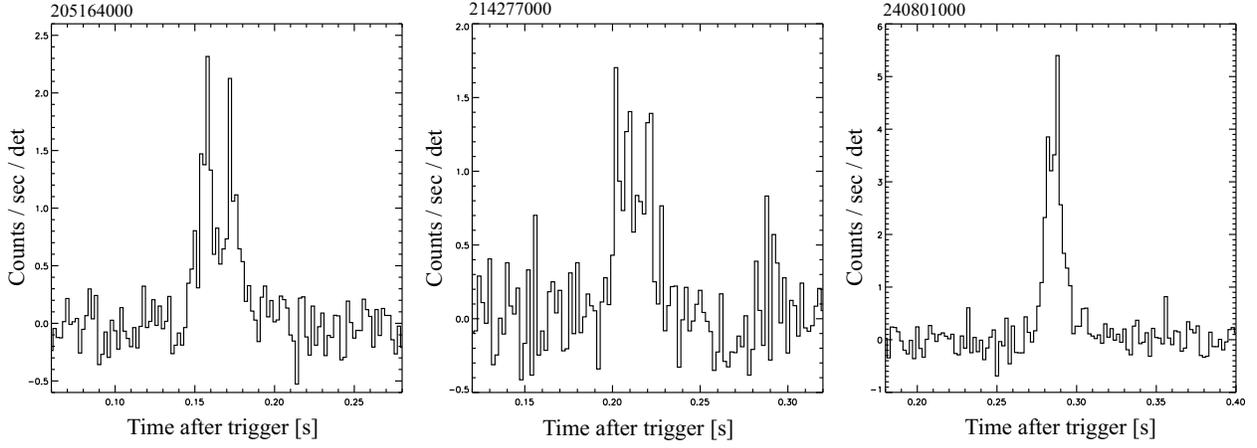}
\caption{15-100\,keV light curves for the short bursts from SGR\,1900$+$14
detected by the {\it Swift} BAT.\label{lc_summary}}
\end{figure}

Prompt follow-up observations by the {\it Swift} XRT were performed
for three short bursts from SGR\,1900$+$14 detected by the {\it Swift} BAT.
They were detected on 2006 April 14 (trigger number 205164),
2006 June 10 (trigger number 214277) and
2006 November 26 (trigger number 240801).
Figure \ref{lc_summary} shows their BAT light curves.
In this paper, we will call these bursts A, B and C, respectively.
Their main properties are summarized in table \ref{trig_summary}.
The $T_{\rm 90}$ durations of these bursts are all $\sim40$\,ms.
Note that the BAT and XRT observations were not simultaneous.

\subsection{Data Reduction}\label{data_reductions}
We used the standard BAT software (HEASoft 6.3.2) and the latest
calibration database (CALDB: 20070924) to process the BAT event data.
The burst pipeline script, {\tt batgrbproduct} (v2.39), was used for
the processing.
Because the bursts were short and weak,
{\tt batgrbproduct} failed to find the burst intervals,
and we used the time interval determined by the flight software for
creating the spectra.
{\tt XSPEC 12.3.1} was used to fit the spectrum \citep{arnaud1996}.
The BAT spectral analyses were performed in the 15-100\,keV band.

The XRT data were systematically analyzed using the
pipeline script.
The cleaned event data from the Window Timing (WT) and
Photon Counting (PC) modes from the {\it Swift}
Science Data Center were used.
Although both WT and PC mode data were processed in the pipeline,
hereafter we focus only on the PC mode data.
The search for the X-ray counterpart, construction
of the X-ray light curve, and fitting the X-ray light
curve and spectra were performed automatically using the standard
XRT softwares and calibration database (HEAsoft 6.3.2 and CALDB: 20070730).
The source region was selected to be a circle of 47$\arcsec$ radius.
The background region was an annulus of outer
radius 150$\arcsec$ and inner radius 70$\arcsec$
excluding the background X-ray sources detected by {\tt ximage}.
Since the count rate in the PC mode was less than 0.1\,counts\,s$^{-1}$, no pile-up
correction was applied in the processing.
The light curve was binned based
on the number of photons required to meet at least a 10$\sigma$
criterion in each bin.
The ancillary response function (ARF) files
were created by {\tt xrtmkarf} (v0.5.5).
The spectra were binned to at least 20 counts in each spectral bin
by {\tt grppha}.
The XRT spectral analyses were performed in the 0.3-10\,keV band.

\subsection{Spectral Analysis}
Since a two blackbody function (2BB) has been suggested as the most
acceptable model for the two SGRs 1806$-$20 and 1900$+$14 using HETE-2
data \citep{nakagawa2007a},
2BB was used for the spectral analysis of the short bursts from SGR\,1900$+$14
detected by the BAT.
If reliable spectral parameters were not obtained,
a single blackbody model (BB) was used instead.
Because we performed the spectral fits using the energy range
above 15\,keV, an absorption model was not applied.

The spectral parameters for the SGR short bursts are summarized
in table \ref{spec_burst_summary}.
The spectra of bursts A and C
are well reproduced by the 2BB model, and their spectral parameters are
consistent with typical values found previously \citep{nakagawa2007a}.
For burst B, BB gives an acceptable result.
The unabsorbed bolometric luminosities
using the spectral parameters for 2BB and BB are
also summarized in table \ref{spec_burst_summary}.

For spectral analysis of the XRT data,
2BB and a blackbody plus a power law model (BB$+$PL) were used.
If reliable spectral parameters were not obtained,
BB and a power law model (PL) were also used.
Since a reliable value of the absorption model was not determined
for some observations, the value was fixed to
$1.91 \times 10^{22}$\,cm$^{-2}$, derived from XMM-Newton
observations \citep{mereghetti2006}.

The results of the XRT spectral analysis
are summarized in table \ref{spec_steady_summary}.
The decreasing trend of burst A is well reproduced by the 2BB and BB$+$PL models.
For bursts B and C, BB and PL give acceptable results.

\subsection{Light Curves}\label{light_curves}
To derive light curves of the
unabsorbed bolometric luminosity,
a conversion factor from count rate to
luminosity was calculated
using the 2BB or BB time-averaged luminosity,
and a time averaged XRT count rate.
In figure \ref{ag_lc_summary}, the X-ray light curves
are shown with the
luminosities of their preceding  short bursts.
Since there are not enough statistics to produce a light curve for burst B,
this event is not considered further.

As seen in figure \ref{ag_lc_summary}, there is a
clear decay in the X-ray emission.
The luminosities at the end of each observation
are 30-50\% lower than the initial X-ray luminosities.
Fitting the light curve with a power law model
($L \propto t^{-p}$, where $L$ is the luminosity,
$t$ is the time since the burst,
and $p$ is the decay index),
the best fit decay index is 0.2$\pm$0.1 for burst A.
Since there are only two data points for burst C,
we calculate the decay index which passes through the two data points
and find 0.4.
The slopes are plotted as dashed lines in figure \ref{ag_lc_summary}.
These decreasing trends can be interpreted as the X-ray afterglows
following the short SGR bursts.

\begin{figure}
\epsscale{1.}
\plotone{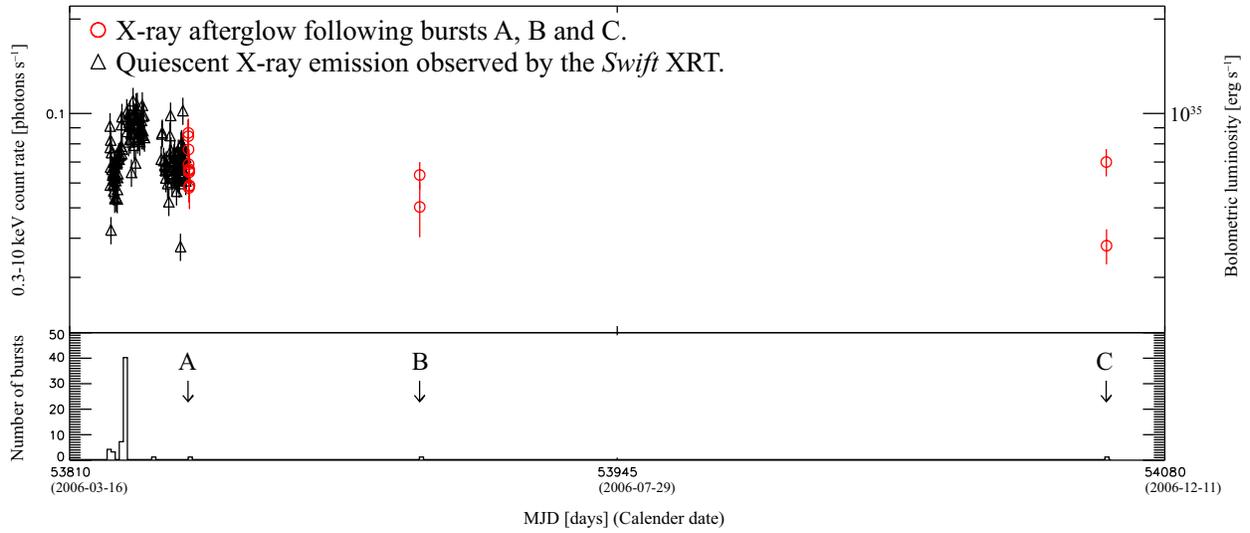}
\caption{Upper panel: Quiescent X-ray emission from SGR\,1900$+$14
observed during {\it Swift} XRT pointed observations (black triangles)
and X-ray afterglow emission (red circles).
Lower panel: Bursting history of SGR\,1900$+$14 in bursts per day.
The three events studied here are indicated by A, B and C.\label{lc_flux_2006_2007}}
\end{figure}

The upper panel in figure \ref{lc_flux_2006_2007} shows
the quiescent X-ray light curve during periods when SGR\,1900$+$14
was emitting short bursts,
while the lower panel presents the burst rates
\footnote{The burst rate is derived from http://www.ssl.berkeley.edu/ipn3/sgrlist.txt.}.
Note the activity at MJD = 53823 days
(2006 March 29); 40 bursts were detected in that one day.
Bursts A, B and C were detected after that active day.
Therefore, we believe that the X-ray emission which we discovered
is not related to the enhanced activity of the SGR.

\section{Discussion and Conclusions}
\begin{figure}
\epsscale{1.}
\plotone{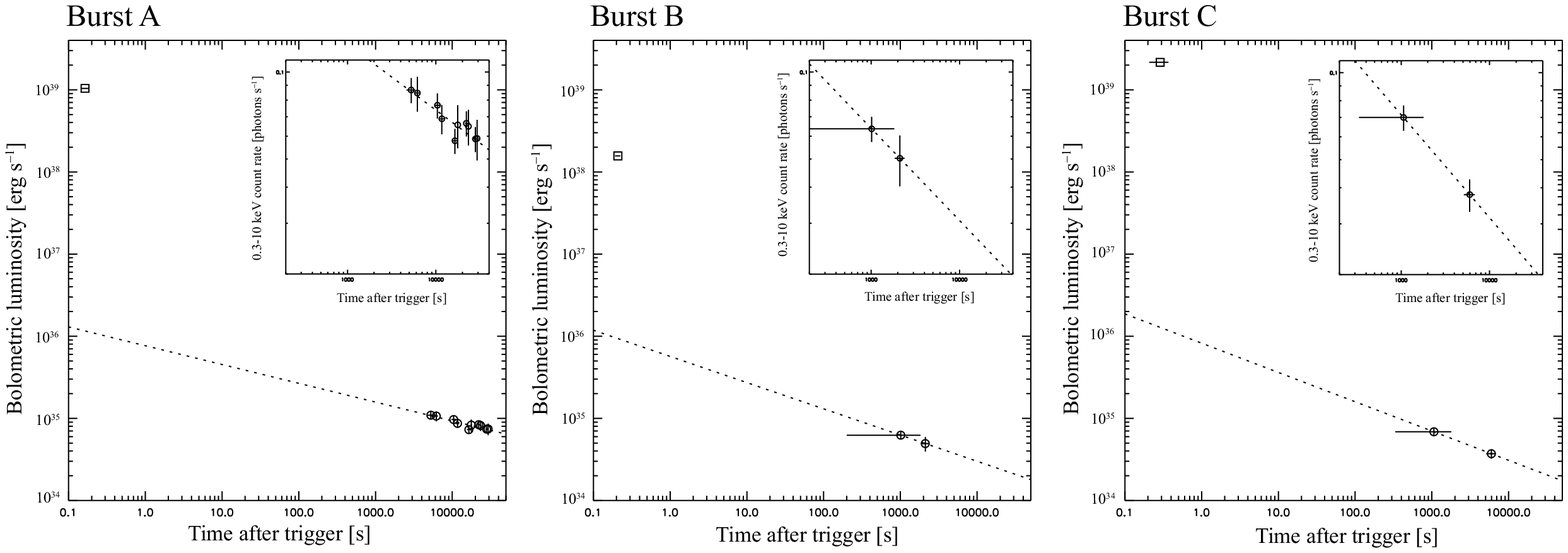}
\caption{Afterglow light curve luminosity for burst A ({\it left}), B ({\it middle}) and C ({\it right}).
The burst luminosities are indicated by the squares.
The dashed lines indicate the extrapolation to the time of the burst.
Inset: light curve in photons, with no spectral correction.\label{ag_lc_summary}}
\end{figure}

As shown in figure \ref{ag_lc_summary},
the luminosities of the SGR short bursts
are two or three orders of magnitude
larger than the backwards-extrapolated values from the SGR X-ray afterglows.
This implies that the decreasing SGR X-ray afterglow is not the tail of
the short bursts.
Therefore this result is different from the X-ray emission accompanying
intermediate bursts from SGR\,1900$+$14,
which is interpreted as the tail of the burst itself \citep{lenters2003}.
Also, compared with the afterglow of the large flare from SGR\,1900$+$14 \citep{feroci2003},
or with the moderate flux
decay of AXP\,CXOU\,J164710.2$-$455216 \citep[e.g.,][]{nakagawa2007b},
the decay time scale of the afterglow accompanying the SGR short bursts
(that is, the time to decrease by a factor of $\sim2$)
is shorter by a factor of $\sim3$.
Since the observing period of the SGR X-ray afterglow ($t < 0.5$\,day)
is different from that of the moderate flux decay of
AXP\,CXOU\,J164710.2$-$455216 ($t \gtrsim 0.5$\,day),
the decay indices and time scales are not exactly comparable.
The initial ($t < 0.5$\,day) decay index $p = 4.8\pm0.5$ of AXP\,1E\,2259$+$586
is much steeper than that of the SGR X-ray afterglow,
despite the fact that the time scale of the
flux decay of AXP\,1E\,2259$+$586 \citep{woods2004}
is consistent with the decay time scale of the SGR X-ray afterglow
(same definition as above).
This suggests that the X-ray afterglows of the short bursts
might be a different phenomenon.
Note that the decay indeces at 0.5 day after the bursts for
the two AXPs CXOU\,J164710.2$-$455216 ($p = -0.3$) and 1E\,2259$+$586 $(p = -0.22)$
are both consistent with the decay index of the SGR X-ray afterglow.

Here, we discuss the analogies and differences
between the X-ray afterglows of short bursts from SGR\,1900$+$14
and the X-ray afterglows of short GRBs.
The decay indices of the SGR X-ray afterglows, $p = $0.2-0.4
(see $\S$ \ref{light_curves} and table \ref{trig_summary}),
are similar to those of the two short GRBs
050724 \citep[$p = 0.6\pm0.2$;][]{campana2006b}
and 051221A \citep[$p = 0.04_{-0.21}^{+0.27}$;][]{burrows2006}.
On the other hand, the five short GRBs
050509B \citep[$p = 1.10_{-0.53}^{+1.26}$;][]{gehrels2005},
050709 \citep[$p \gtrsim 1$;][]{fox2005},
051210 \citep[$p = 2.58\pm0.11$;][]{parola2007},
060313 \citep[$p = 1.46\pm0.08$;][]{roming2006}
and 061201 \citep[$p = 1.90\pm0.15$;][]{stratta2007}
have much steeper temporal indices than those of the SGR X-ray afterglows.
The X-ray afterglow luminosities
of the SGR short bursts are
$L \sim 10^{34}$-$10^{35}$\,erg\,s$^{-1}$.
Considering the cosmological distances to the short GRBs,
the SGR X-ray afterglow luminosity is lower by a factor of
$10^{6}$-$10^{12}$ than the luminosities of the
four short GRBs 050509B, 050709, 050724 and 051221A
\citep[see the following literature for their redshifts;][]{berger2005, prochaska2005, bloom2006, covino2006}.
For the other three GRBs, secure redshift measurements have
not been reported yet.
Thus the indices of the X-ray afterglows of GRBs 050724 and 051221A are
similar to those of the SGR X-ray afterglows,
while the luminosities are different.

Considering the super-Eddington luminosities of the short bursts
($L \gtrsim 10L_{\rm Edd}$ where $L_{\rm Edd} = 1.8 \times 10^{38}$\,erg\,s$^{-1}$ is
the Eddington luminosity with $M = 1.4M_{\sun}$ and $R = 10$\,km),
one possible explanation for the SGR X-ray afterglows might be
a mechanism similar to an external shock in a GRB.
That is, the short burst might be generated by a
relativistic jet from a neutron star, and an interaction
between the jet and an interstellar medium would generate the afterglow.
The lower limit to the number density around SGR\,1900$+$14
$n \gtrsim 0.6(N_{\rm H}/1.91 \times 10^{22}$\,cm$^{-2})(d/3 \times 10^{22}$\,cm$)^{-1}$\,cm$^{-3}$
(where $d$ is the distance to SGR\,1900$+$14)
is reasonable to produce the SGR X-ray afterglow considering
the number density for GRBs \citep{sari1998}.
We can estimate a bulk Lorentz factor assuming a total released
energy $E \sim T_{\rm 90}L \sim 7.6 \times 10^{38}$\,erg for Burst C
and $n \sim 1$\,cm$^{-3}$ \citep{fenimore1996}.
The velocity of the material which is responsible for the external shock
emission might be a weakly relativistic jet with $\gamma \sim 2$ at 1000\,s
after the short burst.
Another possibility might be emission
from the plasma remaining after the short burst,
or an intrinsic quiescent X-ray flux increase.
An alternative simple possibility might be the cooling of the surface
heated by the plasma of the short burst \citep{thompson2002}.
Although some models have been proposed to explain the afterglows
accompanying giant flares \citep{yamazaki2005, lyutikov2006, cea2006},
it is not clear that these models can produce the
afterglow emission from short SGR bursts.
The presence of these afterglows may imply that short burst activity
is in fact much longer than it appears to be.
Indeed, X-ray afterglow emission accompanying short SGR bursts
could be a key to understanding the origin and
emission mechanism of these bursts.
Further prompt multi-wavelength observations of these afterglows
are needed to understand the radiation process.

\acknowledgments

We would like to thank Dr. G. L. Israel for useful comments.
We would like to thank an anonymous referee for comments and suggestions.
YEN is supported by the JSPS Research Fellowships for Young Scientists.
This work is supported in part by a special postdoctoral researchers program in RIKEN.
KH is grateful for support under the Swift Guest Investigator program, NNG04GQ84G.

{\it Facilities:} \facility{Swift (BAT)}, \facility{Swift (XRT)}.

\clearpage


\begin{table}
\begin{center}
\caption{Main properties of three short bursts from SGR\,1900$+$14
observed by {\it Swift}.\label{trig_summary}}
\begin{tabular}{cccccc}
\tableline\tableline
Burst & TrigNum\tablenotemark{a} & Trigger time (UT) & $T_{\rm 90}$ duration\tablenotemark{b} & XRT Exposure Time (PC) & $p$\tablenotemark{c} \\
 &  &  &  (ms) & (ks) & \\
\tableline
A & 205164 & 2006:04:14.04:35:29 & $34\pm9$ & 11.6 & 0.2$\pm$0.1 \\
B & 214277 & 2006:06:10.06:53:01 & $38\pm18$ & 1.9 & \nodata \\
C & 240801 & 2006:11:26.13:16:08 & $40\pm14$ & 1.6 & 0.4 \\
\tableline
\end{tabular}
\tablenotetext{a}{{\it Swift} trigger number.}
\tablenotetext{b}{$T_{\rm 90}$ is the time to accumulate between
5\% and 95\% of the observed photons.}
\tablenotetext{c}{$p$ is the X-ray afterglow decay power law index with 90\% confidence level uncertainties.}
\end{center}
\end{table}

\begin{deluxetable}{clcccccc}
\tabletypesize{\footnotesize}
\tablecaption{
Spectral parameters of three short bursts from SGR\,1900$+$14 observed by the {\it Swift} BAT.\label{spec_burst_summary}
}
\tablewidth{0pt}
\tablehead{
\colhead{Burst} & \colhead{Model} &
\colhead{$kT_{\rm LT}$\tablenotemark{a} or $kT_{\rm BB}$\tablenotemark{b}} &
\colhead{$R_{\rm LT}$\tablenotemark{c} or $R_{\rm BB}$\tablenotemark{b}} & \colhead{$kT_{\rm HT}$\tablenotemark{a}} &
\colhead{$R_{\rm HT}$\tablenotemark{c}} &
\colhead{$L$\tablenotemark{d}} & \colhead{$\chi^2$ (d.o.f.)} \\
\colhead{} & \colhead{} & \colhead{(keV)} & \colhead{(km)} & \colhead{(keV)} & \colhead{(km)} &
\colhead{} & \colhead{}
}
\startdata
A & 2BB & $3.6_{-1.6}^{+2.2}$ & $13_{-9}^{+69}$ & $15_{-3}^{+6}$ & $0.8_{-0.4}^{+0.6}$ & $7.4_{-2.1}^{+10.5}$ & 39 (34)  \\
B & BB & $9.2_{-2.2}^{+2.7}$ & $2.2_{-0.9}^{+1.6}$ & \nodata & \nodata & $4.4_{-1.1}^{+1.0}$ & 18 (16) \\
C & 2BB & $4.9_{-1.2}^{+1.3}$ & $13_{-5}^{+11}$ & $13_{-4}^{+10}$ & $1.3\pm1.0$ & $19.0_{-5.5}^{+4.9}$ & 24 (34) \\
\enddata
\tablenotetext{a}{$kT_{\rm LT}$ and $kT_{\rm HT}$ denote blackbody temperatures for the 2BB fit with 90\% confidence level uncertainties.}
\tablenotetext{b}{$kT_{\rm BB}$ and $R_{\rm BB}$ denote blackbody temperature and radius
for the BB fit with 90\% confidence level uncertainties.}
\tablenotetext{c}{$R_{\rm LT}$ and $R_{\rm HT}$ denote blackbody radii of the 2BB fit with 90\% confidence level uncertainties.}
\tablenotetext{d}{$L$ denotes the bolometric luminosity in units of $10^{39}$\,erg\,s$^{-1}$ with 90\% confidence level uncertainties.}
\end{deluxetable}

\begin{deluxetable}{clcccccccc}
\tabletypesize{\scriptsize}
\tablecaption{
Spectral parameters of observations
after short bursts from SGR\,1900$+$14 by the {\it Swift} XRT.\label{spec_steady_summary}}
\tablewidth{0pt}
\tablehead{
\colhead{Burst} & \colhead{Model} & \colhead{$N_{\rm H}$\tablenotemark{a}} &
\colhead{$kT_{\rm LT}$\tablenotemark{b} or $kT_{\rm BB}$\tablenotemark{c}} &
\colhead{$R_{\rm LT}$\tablenotemark{d} or $R_{\rm BB}$\tablenotemark{c}} & \colhead{$kT_{\rm HT}$\tablenotemark{b}} &
\colhead{$R_{\rm HT}$\tablenotemark{d}} & \colhead{$\Gamma$\tablenotemark{e}} & \colhead{$L$\tablenotemark{f}} & \colhead{$\chi^2$ (d.o.f.)} \\
\colhead{} & \colhead{} & \colhead{} & \colhead{(keV)} & \colhead{(km)} & \colhead{(keV)} & \colhead{(km)} &
\colhead{} & \colhead{}
}
\startdata
A & 2BB & 1.91 & $0.5\pm0.1$ & $3\pm1$ & $1.3_{-0.3}^{+0.6}$ & $0.3\pm0.2$ & \nodata & $1.0\pm0.2$ & 28 (34) \\
       & BB$+$PL & 1.91 & $0.55_{-0.06}^{+0.05}$ & $2.1_{-0.8}^{+0.6}$ & \nodata & \nodata & $1.8_{-1.1}^{+0.5}$ & $5.6_{-4.2}^{+6.2}$ & 31 (35) \\
B & BB & 1.91 & $0.7\pm0.1$ & $1.4_{-0.4}^{+0.5}$ & \nodata & \nodata & \nodata & $0.8\pm0.1$ & 5 (4)  \\
       & PL & 1.91 & \nodata & \nodata & \nodata & \nodata & $2.2\pm0.3$ & $2.4_{-0.8}^{+1.0}$ & 3 (4) \\
C & BB & 1.91 & $0.7\pm0.1$ & $1.5_{-0.5}^{+0.6}$ & \nodata & \nodata & \nodata & $0.6\pm0.1$ & 2 (5) \\
 & PL & 1.91 & \nodata & \nodata & \nodata & \nodata & $2.5_{-0.5}^{+0.6}$ & $1.2_{-0.6}^{+0.9}$ & 5 (5) \\
\enddata
\tablenotetext{a}{$N_{\rm H}$ denotes the absorption. The value of $N_{\rm H}$ is fixed to $1.91 \times 10^{22}$\,cm$^{-2}$
which is derived from XMM-Newton observations \citep{mereghetti2006}.}
\tablenotetext{b}{$kT_{\rm LT}$ and $kT_{\rm HT}$ denote blackbody temperatures of the 2BB fit with 90\% confidence level uncertainties.}
\tablenotetext{c}{$kT_{\rm BB}$ and $R_{\rm BB}$ denote blackbody temperature and radius
for the BB$+$PL or BB fits with 90\% confidence level uncertainties.}
\tablenotetext{d}{$R_{\rm HT}$ and $R_{\rm HT}$ denote blackbody radii of the 2BB fit with 90\% confidence level uncertainties.}
\tablenotetext{e}{$\Gamma$ denotes the power law index of the BB$+$PL or PL fits with 90\% confidence level uncertainties.}
\tablenotetext{f}{$L$ denotes the bolometric luminosity in units of $10^{35}$\,erg\,s$^{-1}$ with 90\% confidence level uncertainties.}
\end{deluxetable}

\clearpage

\end{document}